\documentclass[amsmath,amssymb,superscriptaddress,nobalancelastpage,prb,twocolumn]{revtex4-2}
\usepackage{hyperref}
\usepackage{graphicx}
\usepackage{varioref}
\usepackage{xr-hyper}
\usepackage{xcolor}
\usepackage{nicefrac}
\usepackage{xfrac}
\usepackage{hyperref}
\hypersetup{colorlinks,linkcolor=blue,urlcolor=blue,citecolor=blue}
\usepackage{ulem}
\usepackage{lineno}
\usepackage{amsmath} 
\usepackage{amssymb}
\usepackage{siunitx}
\usepackage{dcolumn}
\usepackage{ulem}
\usepackage{siunitx}
\usepackage{graphicx}
\usepackage{bm}
\usepackage{braket}
\usepackage{wasysym}
\usepackage{textcomp}
\usepackage{pgfplotstable} 

\makeatletter
\def\maketitle{
\@author@finish
\title@column\titleblock@produce
\suppressfloats[t]}
\makeatother

\graphicspath{{Images/}} 

\begin{document} 

\title{Microscopic Origin of Reduced Magnetic Order in a Frustrated Metal}
\date{}

   \author{X. Boraley}
 \affiliation{PSI Center for Neutron and Muon Sciences, 5232 Villigen PSI, Switzerland} 

   \author{O. Stockert}
 \email{oliver.stockert@cpfs.mpg.de}
 \affiliation{Max Plank Institute for Chemical Physics of Solids, DE 01187 Dresden, Germany} 

 \author{J. Lass}
 \affiliation{PSI Center for Neutron and Muon Sciences, 5232 Villigen PSI, Switzerland} 

\author{R. Sibille}
 \affiliation{PSI Center for Neutron and Muon Sciences, 5232 Villigen PSI, Switzerland} 

 \author{\O. S. Fjellv\aa g}
 \affiliation{PSI Center for Neutron and Muon Sciences, 5232 Villigen PSI, Switzerland} 
  \affiliation{ Department for Hydrogen Technology, Institute for Energy Technology, PO Box 40, NO-2027, Kjeller, Norway} 

 \author{S. H. Moody}
 \affiliation{PSI Center for Neutron and Muon Sciences, 5232 Villigen PSI, Switzerland} 

\author{A. M. L\"auchli}
 \affiliation{PSI Center for Scientific Computing, Theory and Data, 5232 Villigen PSI, Switzerland} 
 \affiliation{Institute of Physics, \'Ecole Polytechnique Fédérale de Lausanne (EPFL), 1015 Lausanne, Switzerland} 
 
 \author{V. Fritsch}
 \affiliation{Experimental Physics VI, Center for Electronic Correlations and Magnetism, Institute of Physics, University of Augsburg, 86135 Augsburg, Germany} 
 
\author{D. G. Mazzone}
\email{daniel.mazzone@psi.ch}
\affiliation{PSI Center for Neutron and Muon Sciences, 5232 Villigen PSI, Switzerland}

\begin{abstract}
Although magnetic frustration in metals provides a promising avenue for novel quantum phenomena, their microscopic interpretation is often challenging. Here we use the face-centered cubic intermetallic HoInCu$_4$ as model material to show that Hamiltonians neglecting the charge degree of freedom are appropriate for frustrated metals possessing low density of states at the Fermi surface. Through neutron scattering techniques we determine matching magnetic exchange interactions in the paramagnetic and field-polarized states using an effective spin-1 Heisenberg Hamiltonian, for which we identify antiferromagnetic nearest and next-nearest neighbour interactions $J_1$ and $J_2$ that are close to the critical ratio $J_2$/$J_1$ = 1/2. The study further provides evidence that spin-wave theory fails to predict the low-energy spin dynamics in the antiferromagnetic zero-field state, which is dominated by overdamped magnetic excitations. We conclude that the low-energy fluctuations arise from quantum fluctuations, accounting for the missing moment of the strongly renormalized magnetic long-range order.

\end{abstract}
 \date{\today}
 
   \maketitle


Magnetic frustration arises in materials in which the magnetic interactions cannot be minimized simultaneously due to constraints of the underlying lattice geometry or because of conflicting exchange interactions.  The resulting degeneracy between frustrated ground-state configurations can lead to strong fluctuations triggering unconventional phases of matter, such as (quantum) spin liquids and other exotic spin states  that are of fundamental and technological interest \cite{Balents_2010,Broholm2020,Zhou2017,Vojta_2018,Plumb_2019,Lacroix_2010,Gao2016}. Progress in the microscopic understanding of these exotic magnetic states is frequently achieved through theoretical models, in which Heisenberg, Ising or Kitaev Hamiltonians are used to treat the nearest-neighbour magnetic interactions \cite{Starykh2015,Broholm2020}. These models are particularly successful in describing insulating materials, where further-neighbour interactions decay exponentially with distance and higher-order spin interactions are weak. In contrast, less research has been conducted on itinerant materials \cite{Vojta_2018,Lacroix_2010}, mainly because the dominating contribution of the charge degree of freedom leads to additional challenges. Magnetic exchange interactions in metals typically originate from couplings of the conduction electrons with local moments \cite{Hayami2021,Ruderman_1954,Kasuya_1956,Yoshida_1957}, which can trigger anisotropic exchanges, strong further-neighbour and higher-order exchange interactions. These contributions can either lift the degeneracy of magnetic ground states or trigger novel quantum phases in metals \cite{ Hayami2021,Lacroix_2010,Kurumaji2019,Gao2016} whose interpretation are often challenging. In metals containing Ce, Eu, Sm or Yb, the situation is further complicated, because favorable oxidation states allow for a finite Kondo exchange coupling that screens the local moments \cite{Kondo1964}. This makes it difficult to unambiguously distinguish effects invoked by the Kondo process from magnetic frustration.

Progress in our microscopic understanding of how magnetic frustration is established in itinerant materials can be gained with a stepwise approach, in which we focus first on metals with only minor contributions of the charge degree of freedom and without Kondo screening. In this Letter, the intermetallic compound HoInCu$_{4}$ serves as representative example. Holmium features a stable +3 oxidation state, and the electronic density of states at the Fermi surface in HoInCu$_{4}$ is strongly reduced  \cite{Oliver_2020}. The material crystallizes in a face-centered cubic lattice structure ($a$ = 7.178 \AA) that is prone to frustration \cite{, Chatterji_2019, Matsuura2003,Joshi_2017,Nakamura_1999, Oitmaa2023,Sun2018,Yilidrim_1998,Schick2022,Schick2020, Batalov2016,Kiese2022, Singh2017}. This has been confirmed through a low N\'{e}el temperature $T_\text{N}$ = 0.76 K at which only a fraction of the magnetic entropy is released and  recovered solely at significantly higher temperature \cite{Oliver_2020}. Below $T_\text{N}$ magnetic long-range order is established in a type-III antiferromagnetic (AFM) structure (see Fig. \ref{fig:Zebra_map_DMC_cuts}a) with propagation vector $\vec{q} = (1, \frac{1}{2}, 0)$ in reciprocal lattice units (rlu). Intriguingly, the ordered magnetic moment refined by neutron powder diffraction is significantly lower than the ordered moment expected from the crystal-field ground state triplet \cite{Oliver_2020}. These findings led to the proposal that HoInCu$_{4}$ hosts a partially long-range ordered ground state \cite{Oliver_2020}.

Here we study the microscopic origin of the magnetic ground state. We unambiguously determine the nearest and next-nearest neighbour interactions through neutron diffraction and inelastic neutron scattering in the paramagnetic ($T$ $>$ $T_N$, $\mu_0H$ = 0 T) and field-polarized ($T$ $\sim$ 0 K, $H$ $>$ $H_c$) states, respectively. These results are used to discuss the nature of the antiferromagnetic state, for which we observe overdamped excitations. We find that they arise from quantum fluctuations that lead to moment fluctuations down to lowest temperatures within a magnetic long-range ordered state possessing a strongly reduced ordered moment.


The study was conducted on the same well-characterised HoInCu$_4$ single crystal as in Ref. \cite{Oliver_2020}. The $m$ = 630 mg sample was aligned perpendicular to the $c$-axis, yielding a horizontal ($H$, $K$, 0)-scattering plane in all neutron scattering experiments. The diffraction studies were performed on the thermal and cold diffractometers ZEBRA and DMC at the Paul Scherrer Institut (PSI), Switzerland \cite{Lass2025}. On ZEBRA we used an incoming neutron wavelength of $\lambda$ = 1.383 \AA. The sample was cooled to temperatures below $T$ = 0.1 K employing a cryostat with dilution insert. Reciprocal space maps were taken with sample rotation steps of 0.125 degree, and recorded with a position sensitive area detector. The background was measured in two datasets at $T$ = 50 and 100 K, which were combined for the data analysis. The background-subtracted diffuse neutron scattering signal was analyzed with the Spinteract software package \cite{Paddison_2023}. An incoming wavelength $\lambda$ = 4.5189 \AA~was used for the experiments on DMC, using either dilution or  $^3$He inserts. The zero-field data of these two experiments were combined by accounting for intensity differences and systematic temperature shifts through overlapping data points. The field dependence of the magnetic signal was studied in a vertical 10 T-cyromagnet at fields from $\mu_0H$ = 0 to 5 T with $H||$[0, 0, 1]. In all experiments at DMC, the background was determined at $T$ = 50 K. The inelastic neutron scattering experiment was performed on the multiplexing spectrometer CAMEA at PSI using a dilution insert in a vertical  11 T-cryomagnet to reach a base temperature of $T$ = 40 mK \cite{Camea_2023, Lass_2020}. Excitation spectra were recorded with one degree sample rotation steps with energy transfers $E$ = 0 - 7 meV. We used incident energies $E_i$ = 3.38 and 4.04 meV yielding an elastic resolution of 100 $\mu$eV full width at half maximum (FWHM) to study the low-energy zero-field excitations ($E$ $<$ 0.9 meV) of the material. For these data a background was taken at $\mu_0H$ = 10 T for which the region $E$ $<$ 0.9 meV did not contain any magnetic fluctuations. An $E_i$ = 5 meV was used for the scans under magnetic field, resulting in a FWHM of 180 $\mu$eV and $E$ $<$ 2 meV. Spin-wave calculations were performed through the Su(n)ny and SpinW software packages \cite{Sunny_2023, Toth_2015}. Measurements up to $E$ = 7 meV were recorded to determine the crystal-electric field scheme at zero field and were analyzed with PyCrystalField \cite{PyCristalField_2021}. The refined values from our single crystal inelastic neutron scattering experiment match those obtained from powder samples (see Ref. \cite{Oliver_2020} and Supplemental Material (SM) Note 1). They show that the crystal-field ground state is a triplet, justifying the approximation of an effective spin-1 Hamiltonian at low temperature.


Figures \ref{fig:Zebra_map_DMC_cuts}b and c depict colour-coded single-crystal diffraction maps in the ($H$, $K$, 0)-plane measured on ZEBRA at zero-field and $T$ = 1 and 0.1 K, respectively. Above $T_\text{N}$ = 0.76 K the data reveal a diffuse magnetic signal at reciprocal lattice positions with $\vec{q} = (1, \frac{1}{2}, 0)$, in agreement with the predicted paramagnetic susceptibility of the type-III AFM order \cite{Kiese2022}. At $T$ = 0.1 K $<$ $T_\text{N}$ sharp magnetic Bragg peaks appear at the same wavevectors, but are surrounded by a diffuse signal. The scattering signal was fitted to a combination of two-dimensional Gaussian and Lorentzian lineshapes. Two Gaussian widths account for the axial instrument resolution that leads to an elongated shape of the magnetic Bragg peaks. The diffuse signal was fitted to two-dimensional Lorentzians with a single width, because the inherently broad signal substantially exceeds the instrument resolution. 
\begin{figure}
\centering
\includegraphics[width=1\linewidth,clip]{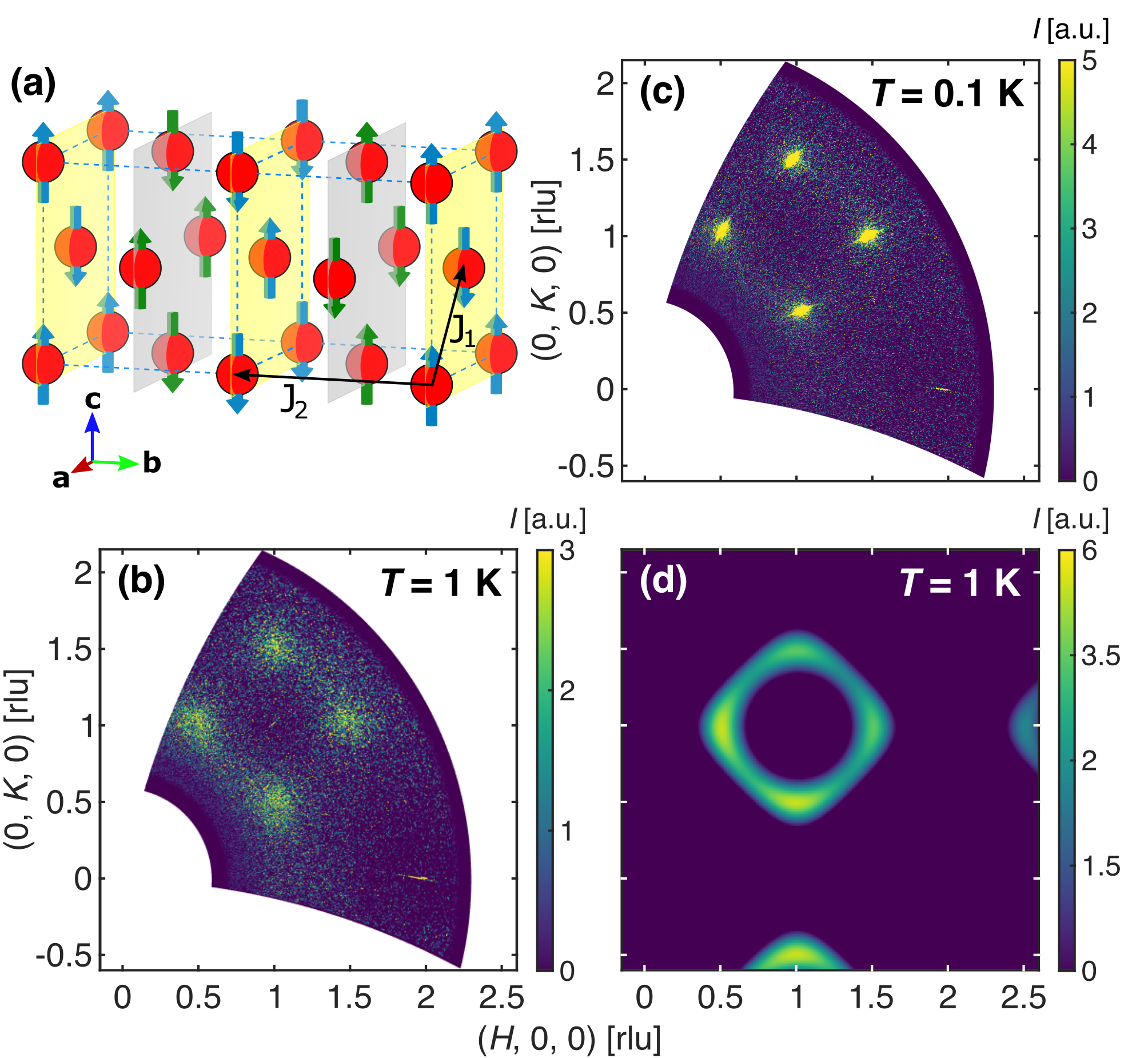}
\caption{\textbf{(a)} Illustration of the type-III antiferromagnetic (AFM) structure. The propagation vector $\vec{q} = (1, \frac{1}{2}, 0)$ in reciprocal lattice units (rlu) splits the face-centered cubic lattice into two independent sublattices, which are indicated with yellow and gray sheets. The magnetic moments corresponding to each sublattice are shown with blue and green arrows. $J_1$ and $J_2$ indicate the magnetic nearest and next-nearest neighbour exchange interactions. \textbf{(b)} and \textbf{(c)} Colour-coded background subtracted diffraction maps of HoInCu$_4$ in the ($H$, $K$, 0)-plane measured at temperatures above and below the N\'{e}el temperature  $T_\text{N}$ = 0.76 K.  The colour bar represents the neutron intensity $I$. \textbf{(d)} Simulated magnetic diffraction pattern at $T$ = 1 K were obtained with the Spinteract software package \cite{Paddison_2023} using the mean of the refined parameters $J_1$ = 0.64(6) K and $J_2$ = 0.29(2) K.}
\label{fig:Zebra_map_DMC_cuts}
\end{figure}

The temperature dependence of the integrated intensities and correlation lengths ($\xi = \frac{a}{\pi \text{FWHM}}$) is presented in Fig. \ref{fig:field_dependence}a and b. The temperature-dependent integrated intensity shows that the magnetic diffuse signal has an onset above $T_\text{N}$  and is continuously increasing as the temperature decreases. This trend is persistent even deep inside the long-range ordered phase. The temperature dependence of the correlation lengths shows that the diffuse short-range order is correlated over roughly 10 unit cells below $T_\text{N}$, while the magnetic long-range order is correlated over 31 unit cells. The latter value is slightly below the instrument resolution (see dashed line in Fig. \ref{fig:field_dependence}b), which likely results from finite magnetic domain sizes. 

\begin{figure}
\centering
\includegraphics[width=1\linewidth,clip]{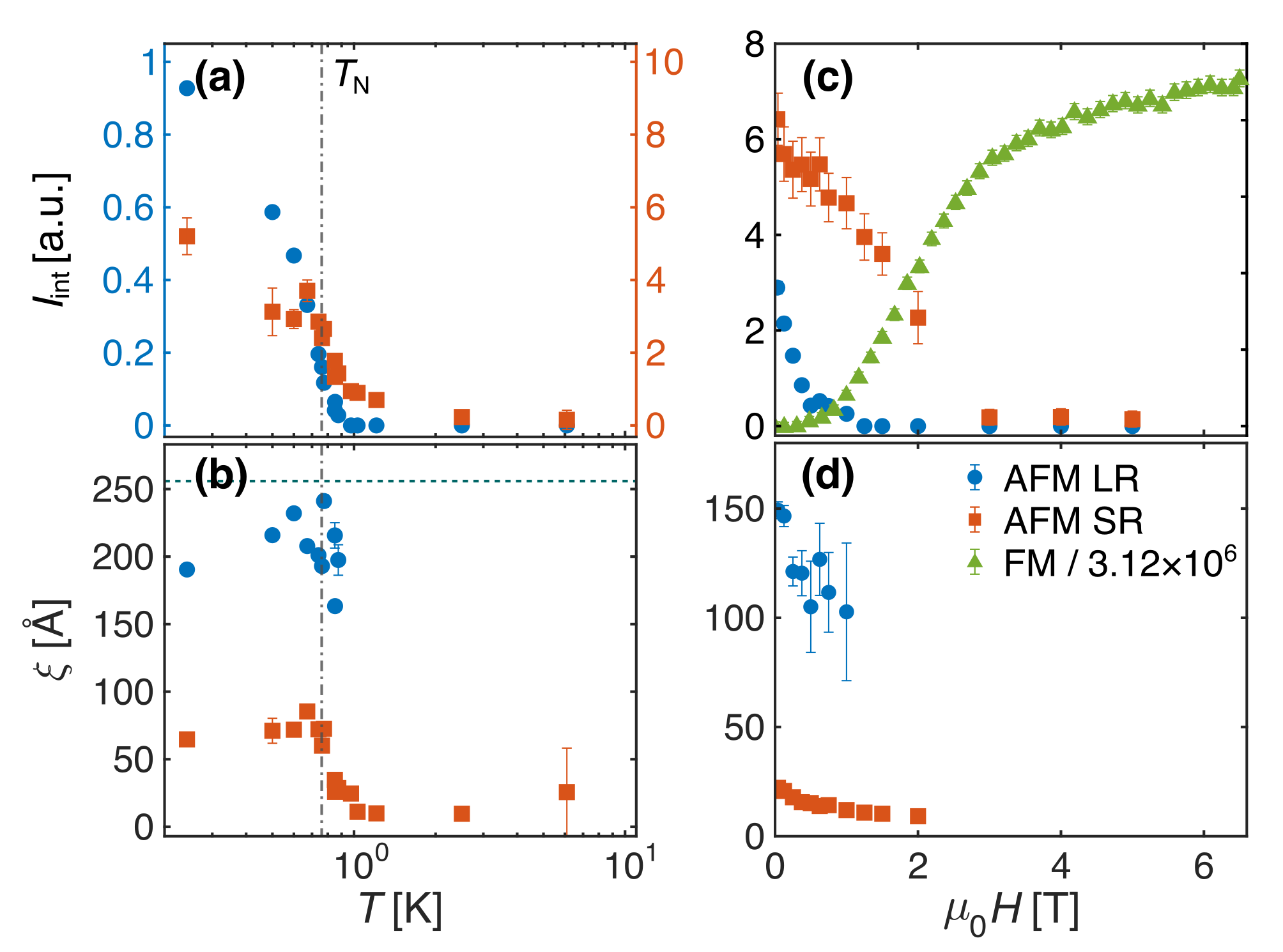}
\caption{ Integrated intensity $I_{int}$ and correlation lengths $\xi = \frac{a}{\pi \text{FWHM}}$ of the AFM long- (LR) and short-range (SR) orders as function of temperature \textbf{(a, b)} and magnetic field \textbf{(c, d)}. The field dependence measured at $T$ = 0.1 K  (measured at DMC) is plotted together with the ferromagnetic (FM) contribution to the structural (2, 0, 0) Bragg peak at  $T$ = 0.04 K (measured at CAMEA).}
\label{fig:field_dependence}
\end{figure}

The magnetic short-range correlations above $T_\text{N}$ hold information on the magnetic exchange interactions in the material. Since the metal possesses a low electronic density of states at the Fermi surface \cite{Oliver_2020}, we approximate the material with a localized system similar to a magnetic insulator. We consider a  nearest ($J_1$) and next-nearest ($J_2$) neighbour Heisenberg Hamiltonian with 
\begin{equation}
\mathcal{H}_{\text{AFM}} = J_1 \sum_{\langle i,j\rangle} \vec{S_i} \cdot \vec{S_j} + J_2 \sum_{\langle \langle i,j\rangle \rangle} \vec{S_i} \cdot \vec{S_j}
\label{equ:Hamiltonian_AFM}
\end{equation}
Here, $\langle i,j\rangle$ and $\langle \langle i,j \rangle \rangle$ uniquely assigns all nearest and next-nearest neighbour bonds, respectively, and $\vec{S_i}$ is a $S$ = 1 spin operator at site $i$.
Our analysis yields antiferromagnetic exchange interactions $J_1$ = 0.64(6) K and $J_2$ = 0.29(2) K with $\chi_{red}^2$ = 0.25 (see Fig. \ref{fig:Zebra_map_DMC_cuts}d and SM Note 2 for further details). The resulting $J_2$/$J_1$ = 0.45(5) ratio is compatible with the observed type-III AFM order predicted for 0 $<$ $J_2$/$J_1$ $<$ 0.5 \cite{Sun2018, Oitmaa2023}. Notably, our results classify HoInCu$_4$ as a rare type-III AFM case close to the type-II AFM phase boundary with $\vec{q} = (\frac{1}{2}, \frac{1}{2}, \frac{1}{2})$ predicted for $J_2$/$J_1$ $\geq$ 0.5 \cite{Sun2018, Oitmaa2023, Joshi_2017}.

Figures \ref{fig:field_dependence}c and d show the magnetic field dependence of the magnetic long- and short-range orders for $H||$[0, 0, 1] and $T$ = 0.1 K. The results stem from one-dimensional Gaussian and Lorentzian fits to reciprocal space maps recorded on DMC. We observe that the integrated magnetic Bragg intensity corresponding to the AFM long-range order is suppressed above $\mu_0H \approx 1$ T. In contrast the short-range correlations survive to larger magnetic fields of $\mu_0H \approx 2.5$ T, revealing a field range between $\mu_0H \approx$ 1 and 2.5 T in which only short-range correlations are present. The magnetization of the material was probed via the ferromagnetic (FM) contribution to the structural (2, 0, 0) Bragg peak intensity at $T$ = 40 mK on CAMEA, supporting that the field-polarized state is reached above $\mu_0H$ $>$ $\mu_0H_c$ = 2.5 T.


The field-polarized state serves as an alternative way to determine the magnetic interaction parameters, as in this regime linear spin-wave theory is known to be exact. Figure \ref{fig:FM_spin-wave}a shows the field-polarized excitation spectrum at $\mu_0H = 5$ T along a representative path in the ($H$, $K$, 0)-plane. The spectrum was measured at $T$ = 40 mK on CAMEA. The field-polarized excitation spectra were analyzed using the same spin-1 Hamiltonian as in Eq. \ref{equ:Hamiltonian_AFM} with an additional Zeeman term accounting for the magnetic field anisotropy.
\begin{figure}
\centering
\includegraphics[width=1\linewidth,clip]{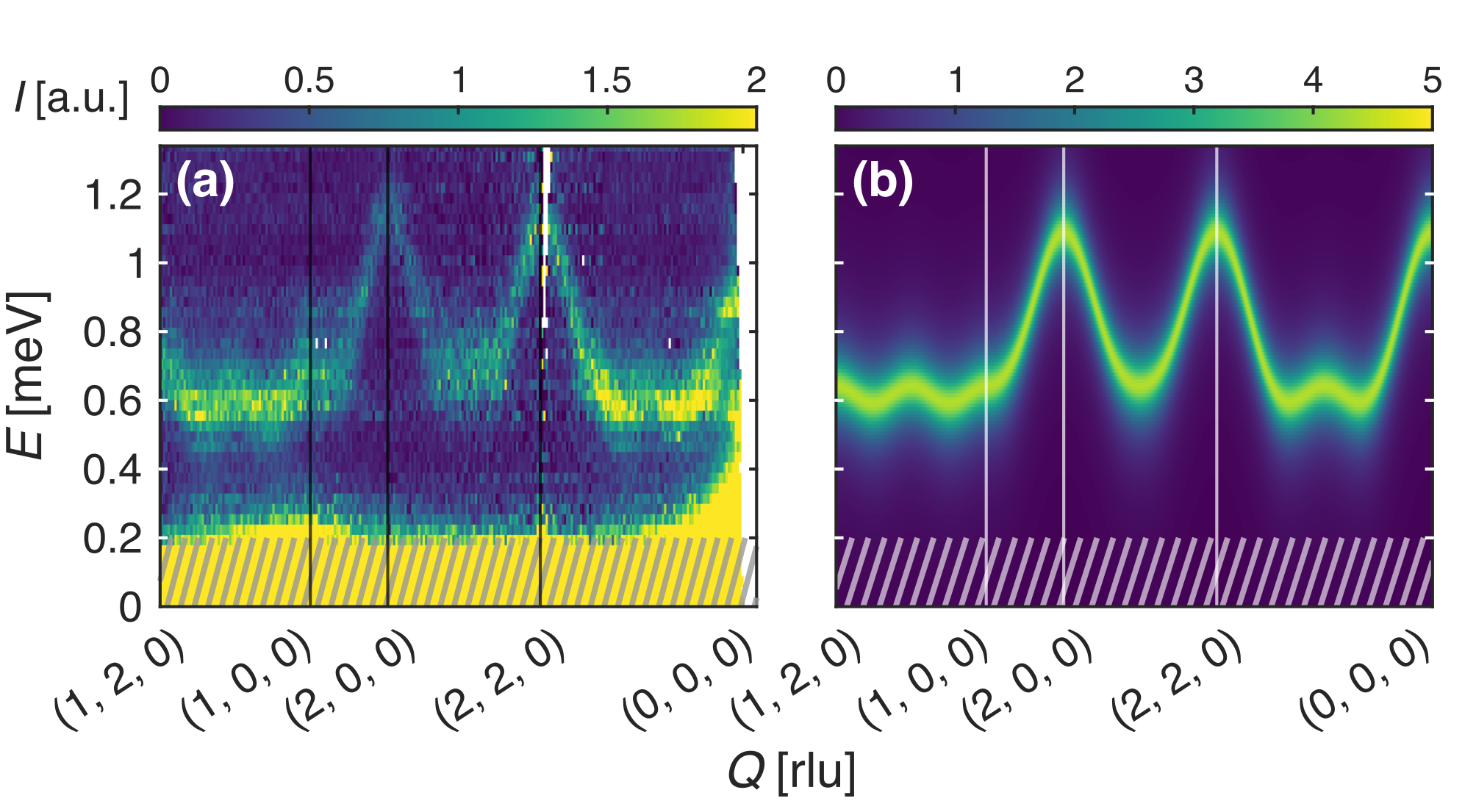}
\caption{\textbf{(a)} Spin-wave spectrum of HoInCu$_4$ at $\mu_0H$ = 5 T and $T$ = 40 mK for a representative path through the ($H$, $K$, 0)-plane. An integration width of $\Delta Q$ = 0.35 rlu perpendicular to the cut was used. \textbf{(b)} Linear spin-wave simulation obtained with the SpinW software package \cite{Toth_2015} using the mean of the refined parameters $J_1$ = 0.66(3) K and $J_2$ = 0.30(3) K. A Lorentzian smearing $\Delta E$ = 0.08 meV half width half maximum was used to mimic the instrument resolution. The hashed region at low energy transfers $E$ represents the elastic line resolution.}
\label{fig:FM_spin-wave}
\end{figure}
\begin{equation}
\mathcal{H_{\text{FP}}} = \mathcal{H_{\text{AFM}}} - g\mu_B\mu_0H \sum_{i} S_i^{z}
\label{equ:Hamiltonian_FM}
\end{equation}

The effective $g$-factors at $\mu_0H$ = 4, 5 and 6.5 T were determined from the polarized spin-wave maxima and compared with the predicted field evolution of the crystal-field ground state triplet. We found that the linear Zeeman term is only approximate for the crystal-field scheme of HoInCu$_4$. Thus, a field dependent g-factor was assumed for the effective spin-1 treatment, and note that more accurate results can be obtained if all crystal-field wave-functions and dipole-dipole interactions are considered in the spin-wave analysis  (see SM Note 3 and 6 for details). Within the effective spin-1 treatment refinements over 820 data points yield $J_1$ = 0.66(3) K and $J_2$ = 0.30(3) K with $\chi_{red}^2$ = 0.39, which is in  agreement with our analysis of the magnetic correlations above $T_\text{N}$. The corresponding linear spin-wave simulation at $\mu_0H$ = 5 T  is plotted in Fig. \ref{fig:FM_spin-wave}b.

\begin{figure}
\centering
\includegraphics[width=\linewidth,clip]{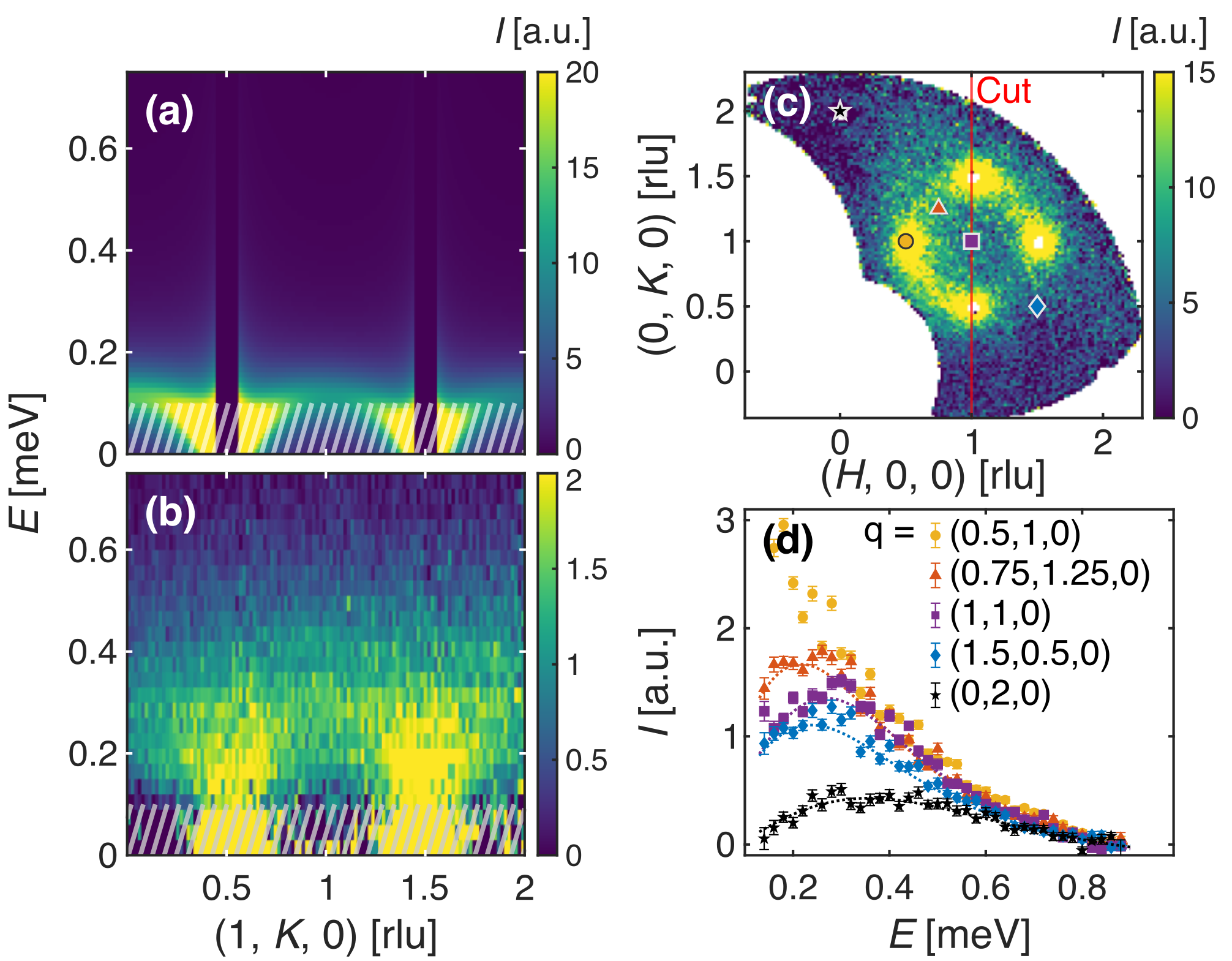}
\caption{\textbf{(a)} Linear spin-wave simulation of the type-III AFM state along (1, $K$, 0). The divergence occurring at the magnetic positions has been removed for clarity. The instrument resolution was approximated with a Lorentzian smearing $\Delta E$ = 0.04 meV half width half maximum. \textbf{(b)} Background subtracted magnetic excitation spectrum along  (1, $K$, 0), using an integration width $\Delta Q$ = 0.35 rlu perpendicular to the cut \textbf{(c)} Reciprocal space map at an energy transfer $E$ = 0.00(5) meV. The intensity at the magnetic magnetic Bragg peaks have been removed for clarity. The cut indicated in the panel represents the path plotted in panel \textbf{(b)}. \textbf{(d)} Energy cuts at the reciprocal space positions indicated in panel \textbf{(c)}, with circular integration width $\Delta Q$ = 0.2 rlu. The data are overplotted with damped harmonic oscillator fits, revealing a relaxation rate of  $\sim$4$J_1$.}
\label{fig:AFM_excitations}
\end{figure}

Having determined matching interaction parameters in the paramagnetic and field-polarized states, we expect these parameters will also be applicable in the type-III AFM ground state. Using linear spin-wave theory magnetic excitations are expected below an energy transfer $E$ = 0.2 meV (see Fig. \ref{fig:AFM_excitations}a). The observed magnetic excitation spectrum at $T$ = 40 mK is displayed in Fig. \ref{fig:AFM_excitations}b (see also SM Note 4 for further cuts and simulations). Their comparison provides evidence that linear spin-wave theory fails to capture the low-energy spin dynamics in HoInCu$_4$. In contrast to the predictions we observe column-like excitations centered around the magnetic Bragg peak positions. The excitations are most intense around $E$ $\approx$ 0.3 meV and display only a weak momentum dependence. The life-time of the damped excitations was assessed through one-dimensional energy cuts at specific reciprocal lattice positions indicated in Fig. \ref{fig:AFM_excitations}c, showing a thin slice ($E$ = 0.00(5) meV) around the elastic ($H$, $K$, 0)-plane. The resulting cuts shown in Fig. \ref{fig:AFM_excitations}d were fitted to a damped harmonic oscillator, for which we find a global relaxation rate $\Gamma$ = 0.24(2) meV (2.8(2) K) corresponding to roughly $\sim$4$J_1$. 

Figure \ref{fig:AFM_excitations}c reveals a broad elastic intensity distribution that is centered around the magnetic Bragg peaks and shows some finite intensity along the shortest path connecting neighbouring peaks. This result provides evidence that within our resolution of 0.1 meV, a substantial fraction of the Ho moments remain short-range correlated. We estimated the fluctuating moment fraction through a refinement of the diffuse scattering signal that was measured below $T_{\text{N}}$ during the diffraction experiment on ZEBRA (see Fig. \ref{fig:Zebra_map_DMC_cuts}c), using the interaction parameters obtained from the analysis of the paramagnetic state (see SM Note 2 for details). The results are in agreement with the experimental observations, if about 30\%  of the Ho moment remains fluctuating within the long-range ordered state. 

A strong renormalization of the ordered magnetic moment together with complex dynamic many-body properties is a trait mark of quantum effects in frustrated magnets \cite{Balents_2010,Broholm2020,Zhou2017,Li2022,Ferrari2018,Oitmaa2023,Sun2018,Yilidrim_1998,Schick2022,Schick2020}. While frustrated triangular and square lattices are prime geometries for quantum effects to appear \cite{Li2022,Ferrari2018}, an increasing amount of theoretical studies have found that quantum fluctuations are also important in face-centered cubic lattices \cite{Oitmaa2023,Sun2018,Yilidrim_1998,Schick2022,Schick2020}. For instance, it is predicted that the type-III AFM state is unstable at the phase boundaries $J_2/J_1$ = 0 and 0.5, but that thermal fluctuations can help to stabilize the structure \cite{Sun2018,Oitmaa2023,Schick2022}. While such an order-by-disorder transition has been recently observed in K$_2$IrCl$_6$  close to  $J_2/J_1$ = 0 \cite{Wang2024}, the inaccessibility of materials close to $J_2/J_1$ = 0.5 hitherto hindered experimental studies close to the type-II/type-III AFM phase boundary. HoInCu$_4$ bridges this issue, because our results provide compelling evidence that the material can be modeled as an effective spin-1 Heisenberg system with $J_2$/$J_1$ = 0.45(5). The importance of quantum fluctuations was assessed using spin-wave theory including quantum corrections of the leading term in 1/$S$. We obtain that the expected ordered moment of HoInCu$_4$ is renormalized by about 30\%, which matches the estimated fluctuating moment fraction. These results are also in agreement with previous results reporting a mismatch between the ordered moment expected from the crystal-field ground state triplet ($\mu_{CEF}$ = 4.58$\mu_B$), and the magnetic moment ($\mu$ = 3.23(4)$\mu_B$) refined from neutron powder diffraction \cite{Oliver_2020}. Thus, our results provide evidence for the presence of substantial quantum fluctuations, accounting for the missing moment of the renormalized magnetic long-range order.

We note that neutron scattering can apriori not discriminate a case in which the ordered moment is homogeneously reduced, from a spatially inhomogenous scenario in which half of the Ho ions carry the full moment and the other half remain disordered (see also SM Note 4). The latter case is enabled through the ordering wavevector $\vec{q} = (1, \frac{1}{2}, 0)$, splitting the face-centered cubic type-III AFM structure into two orbits with independent Ho sublattices (see Fig. \ref{fig:Zebra_map_DMC_cuts}a) \cite{Anderson1950,Villain1959,Nakamura_1999,Oliver_2020}. While modern theoretical considerations of frustrated magnets favor homogeneous solutions \cite{Li2022,Ferrari2018,Oitmaa2023}, local probes such as muon spin rotation and nuclear spin resonance or neutron diffraction under uniaxial pressure may be able to experimentally distinguish the two scenarios. We further mention that HoInCu$_4$ potentially hosts an intriguing quantum phase between $\mu_0H \approx$ 1 and 2.5 T, at which magnetic long-range order is suppressed but short-range correlations are still present. In fact, linear spin-wave theory predicts the critical field around $\mu_0H_c \approx$ 2.5 T (see SM Note 5), but fails to capture the nature of the magnetic state in this field range. Future theoretical and experimental efforts are required to elucidate the microscopic nature of this phase. 

Finally, we claim that the reported study has an impact beyond the field of frustrated magnetism in face-centered cubic lattices. Although magnetic frustration is often challenging to interpret in metals, we here show that progress in the microscopic understanding of frustrated Fermi liquids can be gained with a step-wise approach. We elucidate that local Hamiltonians neglecting the charge degree of freedom provide an accurate description of frustrated magnetic properties in metals possessing low-density of states at the Fermi surface. We mention that for frustrated face-centered cubic metals Hubbard models have shown a promising way forward to appropriately account for itinerant effects if they become relevant \cite{Singh2017} Thus, we hypothesize that further such experimental and theoretical efforts can build a foundation to gain insight in how magnetic frustration is established in metals, in which the charge degree of freedom plays a significant role.


In summary, we have studied the microscopic origin of the magnetic ground state in metallic HoInCu$_4$ possessing only minor contribution of the charge degree of freedom at the Fermi level.  Using an effective spin-1 Heisenberg Hamiltonian we find matching nearest and next-nearest neighbour exchange couplings in the paramagnetic and magnetic field-polarized states. The resulting $J_2$/$J_1$ = 0.45(5) ratio is compatible with theoretical models for face-centered cubic lattices, identifying the material as a rare type-III AFM close to the phase boundary $J_2$/$J_1$ $=$ 0.5 \cite{Sun2018, Oitmaa2023,Joshi_2017}. In contrast to classical predictions we find a non-trivial type-III AFM state with an overdamped magnon spectrum below $T_N$ and at zero magnetic field. Our investigations provide evidence that the low-energy fluctuations arise from quantum fluctuations, accounting for the missing moment of the renormalized ordered moment value in the AFM ground state. This motivates experimental and theoretical studies in face-centered cubic lattices close to the $J_2$/$J_1$ $=$ 0.5 phase boundary.

We thank Joseph Paddison and Oksana Zaharko for fruitful discussions. We acknowledge the Paul Scherrer Institut for the allocated beam time on ZEBRA, DMC and CAMEA, alongside the Swiss National Foundation for financial support (Grant No. 200021\_200653). The experimental data used in this work can be found using the link \href{https://doi.org/10.5281/zenodo.14864167}{https://doi.org/10.5281/zenodo.14864167}

\bibliography{sample}


\newpage

\newcommand{\beginsupplement}{
        \setcounter{table}{0}
        \renewcommand{\thetable}{S\arabic{table}}
        \setcounter{figure}{0}
        \renewcommand{\figurename}{\textbf{Supplemental Figure}}}

\graphicspath{{./}{SI/}}
\newcommand{\RM}[1]{\MakeUppercase{\romannumeral #1{}}}

\beginsupplement

\newcommand{\tcr}[1]{\textcolor{black}{#1}}

\title{Supplemental Material for: Microscopic Origin of Reduced Magnetic Order in a Frustrated Metal}
\date{\today}

\begin{abstract}
\end{abstract}

\pacs{}
\maketitle
\onecolumngrid{

\vspace{-10ex}
\subsection*{Supplemental Note 1. Crystal Electric Field Scheme}
The crystal-electric field (CEF) scheme of HoInCu$_4$ was determined at $T$ = 6.5 and 50 K using CAMEA, PSI  equipped with a 11 T-vertical cryo-magnet with dilution insert. The crystal was aligned in the horizontal ($H$, $K$,0)-scattering plane, using GE-Varnish that was covered with a thin cadmium sheet to minimize parasitic scattering. In each scan the sample was rotated in 1 degree steps over  100 degrees, covering the structural (2, 0, 0) and (0, 2, 0) Bragg peaks up to $|\vec{Q}| = 2.24~$\AA$^{-1}$.  For the $T$ = 6.5 K measurement we used incoming energies $E_i$ = [5, 6, 7, 8, 9, 10] meV at detector positions $2\theta$ = [-54, -50, -47, -44, -42, -40] degrees. Each setup consisted of four individual scans in between which the detector position was shifted by 4 degrees and/or the energy by 0.13 meV to cover all blind spots between the analyzer units. For each scan  a monitor $M$ = 100'000 was employed, which corresponds to $\tau$ = [23, 25, 25, 26, 28, 34] sec/point. At $T$ = 50 K we used $M$ = [1'000'000, 62'500, 62'500, 62'500, 62'500] corresponding to $\tau$ =[225, 16, 16, 18, 21] sec/point. The data reduction was performed with the MJOLNIR software package \cite{Lass_2020} (version 1.3.1), with which we also masked the Currat-Axe spurions with circles of $\Delta Q$ = 0.07 \AA$^{-1}$~in diameter.  Supplemental (Suppl.) Figure \ref{fig:CEF_fit} shows the resulting excitation spectrum integrated from $|\vec{Q}|$ = 1 to 2 \AA$^{-1}$~with an energy step size of $\delta E$ = 0.075 meV ranging from energy transfers $E$ = 0.2 to 7 meV. The plots reveal excitations at energy transfers $E\approx$ 1.35, 1.70, 2.62, 4.14 4.63 and 6.34 meV.

\begin{figure}
\centering
\includegraphics[width=0.5\linewidth,clip]{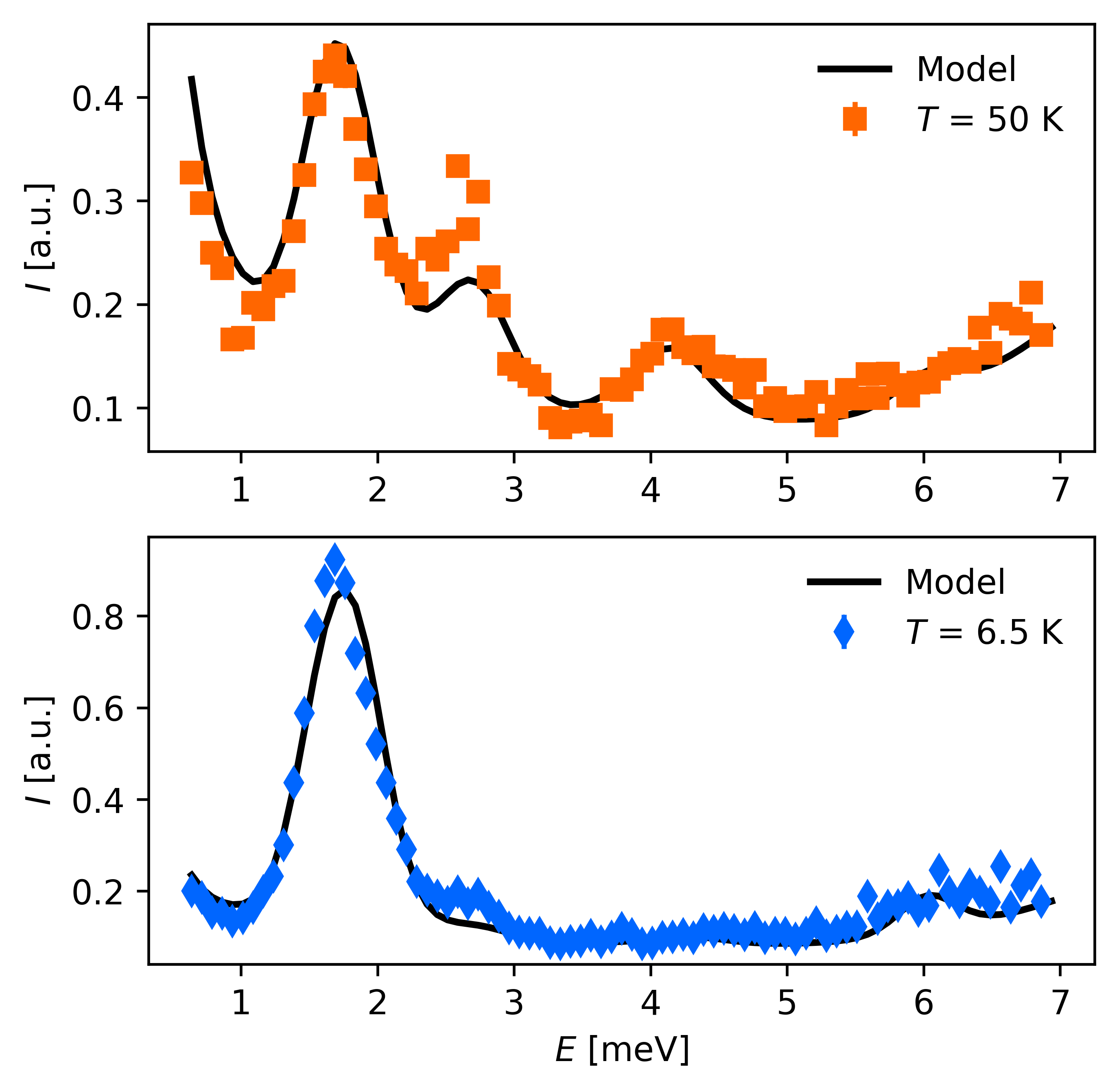}
\caption{Integrated intensity over |$Q$| = 1 to 2 \AA$^{-1}$~as function of energy transfer $E$ at $T$ = 50 and 6.5 K. The data points are overplotted with the intensity expected from the determined CEF model (black line).}
\label{fig:CEF_fit}
\end{figure}

The CEF refinement was performed with PyCrystalField \cite{PyCristalField_2021} (version 2.3.1). The cubic symmetry of HoInCu$_4$ results in a CEF Hamiltonian of the form \cite{Sunny_2023}

\begin{equation}
H_{CEF} = B_{4} (O_{4}^{0} + 5O_{4}^{4}) + B_{6} (O_{6}^{0} - 21O_{6}^{4}).
\label{equ:Hamiltonian_CEF}
\end{equation}

Here, $O_{l}^{m}$ are the Stevens operators and  $B_{l}$ the CEF parameters. The background of the neutron spectra was modeled with two Gaussians and a constant offset to account for incoherent scattering of the elastic line, a constant background noise and increased parasitic signal above $E$ = 6 meV, originating from the missing overlapping measurement at $E_i$ = 11 meV. The parameters amount to ($I_{int1}$, $cen_{1}$, $\sigma_{1}$, $I_{int2}$, $cen_{2}$, $\sigma_{2}$, const) = (0.4, 0.0.7, 0.42, 8, 1, 0.08) and (0.15, 0.0.7, 0.42, 8, 1, 0.08) for $T$ = 50 and 6.5 K, respectively. An iterative refineement of the CEF parameters using a constant energy resolution across $E$ and values from Ref. \cite{Oliver_2020} as starting parameters yield the global solution $B_{4}$ = -0.2701 
and $B_{6}$  = 0.0064 mK. The resulting Eigenvalues and Eigenvectors are shown in Table S1, and the spectrum is overplotted with the experimental results in Suppl. Fig. \ref{fig:CEF_fit}. Consistent with Ref. \cite{Oliver_2020} we find a ground-state triplet that is separated from the first excited doublet by $E\approx$ 1.6 meV. This justifies the approximation of an effective spin-1 Hamiltonian at low temperature. We mention that since the CEF wavefunctions of the ground state triplet and excited doublet are orthogonal, we expect no hybridization of the ground state exctiations with the excited CEF state.

\begin{table}

\begin{ruledtabular}
\begin{tabular}{c|ccccccccccccccccc}
E [meV] &$|-8\rangle$ & $|-7\rangle$ & $|-6\rangle$ & $|-5\rangle$ & $|-4\rangle$ & $|-3\rangle$ & $|-2\rangle$ & $|-1\rangle$ & $|0\rangle$ & $|1\rangle$ & $|2\rangle$ & $|3\rangle$ & $|4\rangle$ & $|5\rangle$ & $|6\rangle$ & $|7\rangle$ & $|8\rangle$ \tabularnewline
 \hline 
0.000 & 0.0 & -0.722 & 0.0 & 0.0 & 0.0 & -0.581 & 0.0 & 0.0 & 0.0 & 0.367 & 0.0 & 0.0 & 0.0 & 0.085 & 0.0 & 0.0 & 0.0 \tabularnewline
0.000 & 0.0 & 0.0 & 0.0 & -0.085 & 0.0 & 0.0 & 0.0 & -0.367 & 0.0 & 0.0 & 0.0 & 0.581 & 0.0 & 0.0 & 0.0 & 0.722 & 0.0 \tabularnewline
0.000 & 0.0 & 0.0 & -0.401 & 0.0 & 0.0 & 0.0 & -0.583 & 0.0 & 0.0 & 0.0 & 0.583 & 0.0 & 0.0 & 0.0 & 0.401 & 0.0 & 0.0 \tabularnewline
1.614 & 0.0 & 0.0 & -0.588 & 0.0 & 0.0 & 0.0 & -0.393 & 0.0 & 0.0 & 0.0 & -0.393 & 0.0 & 0.0 & 0.0 & -0.588 & 0.0 & 0.0 \tabularnewline
1.614 & -0.375 & 0.0 & 0.0 & 0.0 & -0.347 & 0.0 & 0.0 & 0.0 & 0.691 & 0.0 & 0.0 & 0.0 & -0.347 & 0.0 & 0.0 & 0.0 & -0.375 \tabularnewline
1.877 & 0.553 & 0.0 & 0.0 & 0.0 & 0.441 & 0.0 & 0.0 & 0.0 & 0.0 & 0.0 & 0.0 & 0.0 & -0.441 & 0.0 & 0.0 & 0.0 & -0.553 \tabularnewline
1.877 & 0.0 & 0.440 & 0.0 & 0.0 & 0.0 & 0.027 & 0.0 & 0.0 & 0.0 & 0.824 & 0.0 & 0.0 & 0.0 & 0.355 & 0.0 & 0.0 & 0.0 \tabularnewline
1.877 & 0.0 & 0.0 & 0.0 & 0.355 & 0.0 & 0.0 & 0.0 & 0.824 & 0.0 & 0.0 & 0.0 & 0.027 & 0.0 & 0.0 & 0.0 & 0.440 & 0.0 \tabularnewline
2.161 & -0.411 & 0.0 & 0.0 & 0.0 & -0.270 & 0.0 & 0.0 & 0.0 & -0.718 & 0.0 & 0.0 & 0.0 & -0.270 & 0.0 & 0.0 & 0.0 & -0.411 \tabularnewline
4.392 & 0.0 & 0.0 & -0.583 & 0.0 & 0.0 & 0.0 & 0.401 & 0.0 & 0.0 & 0.0 & -0.401 & 0.0 & 0.0 & 0.0 & 0.583 & 0.0 & 0.0 \tabularnewline
4.392 & 0.0 & -0.185 & 0.0 & 0.0 & 0.0 & 0.173 & 0.0 & 0.0 & 0.0 & -0.303 & 0.0 & 0.0 & 0.0 & 0.919 & 0.0 & 0.0 & 0.0 \tabularnewline
4.392 & 0.0 & 0.0 & 0.0 & 0.919 & 0.0 & 0.0 & 0.0 & -0.303 & 0.0 & 0.0 & 0.0 & 0.173 & 0.0 & 0.0 & 0.0 & -0.185 & 0.0 \tabularnewline
6.027 & 0.441 & 0.0 & 0.0 & 0.0 & -0.553 & 0.0 & 0.0 & 0.0 & 0.0 & 0.0 & 0.0 & 0.0 & 0.553 & 0.0 & 0.0 & 0.0 & -0.441 \tabularnewline
6.027 & 0.0 & 0.0 & 0.0 & -0.149 & 0.0 & 0.0 & 0.0 & 0.306 & 0.0 & 0.0 & 0.0 & 0.795 & 0.0 & 0.0 & 0.0 & -0.502 & 0.0 \tabularnewline
6.027 & 0.0 & -0.502 & 0.0 & 0.0 & 0.0 & 0.795 & 0.0 & 0.0 & 0.0 & 0.306 & 0.0 & 0.0 & 0.0 & -0.149 & 0.0 & 0.0 & 0.0 \tabularnewline
6.055 & 0.0 & 0.0 & -0.393 & 0.0 & 0.0 & 0.0 & 0.588 & 0.0 & 0.0 & 0.0 & 0.588 & 0.0 & 0.0 & 0.0 & -0.393 & 0.0 & 0.0 \tabularnewline
6.055 & 0.436 & 0.0 & 0.0 & 0.0 & -0.554 & 0.0 & 0.0 & 0.0 & -0.083 & 0.0 & 0.0 & 0.0 & -0.554 & 0.0 & 0.0 & 0.0 & 0.436 \tabularnewline
\end{tabular}\end{ruledtabular}
\label{tab:CEF_eigenvalues}
\caption{Eigenvalues and Eigenvectors obtained from the CEF refinements at $T$ = 6 and 50 K.}
\end{table}

\subsection*{Supplemental Note 2. Analysis of Diffuse Scattering Results}
The analysis of the paramagnetic signal was studied on the thermal diffractometer ZEBRA, PSI using a position-sensitive area detector. A cryostat with dilution insert was used to record data with a coverage of 80 degrees in sample rotation to map at least one full Brillouin zone (4 magnetic and one structural Bragg peaks). We used a step size of 0.125 degrees and $\lambda$ = 1.383 \AA~incoming neutrons. Each point was recorded at $2\theta$ = 16.675 degrees with 250'000 monitor, which roughly corresponds to 37.5 sec/point. $q$-maps were recorded at $T$ = [0.1, 0.7, 0.8, 0.85, 0.9, 0.95, 1, 4, 10] K and a combined background at $T$ = 50 and 100 K was subtracted from the data.

The nearest ($J_1$) and next-nearest ($J_2$) interaction parameters of a spin-1 Heisenberg Hamiltonian

\begin{equation}
H = J_1 \sum_{\langle i,j\rangle} \vec{S_i} \cdot \vec{S_j} + J_2 \sum_{\langle \langle i,j\rangle \rangle} \vec{S_i} \cdot \vec{S_j}
\label{equ:Hamiltonian_Spinteract}
\end{equation}

were determined with the Spinteract software \cite{Paddison_2023} (version as of 14 September 2023).  $\langle i,j\rangle$ and $\langle \langle i,j\rangle \rangle$ uniquely assigns the nearest and next-nearest neighbours, respectively and $\vec{S_i}$ is a spin-1 operator on site $i$. Initially, only datasets above the N\'eel temperature were considered. The $T$ = 1 K data with strongest diffuse signal was refined first to determine the scaling factor. The obtained factor was fixed in subsequent refinements, leaving only $J_1$ and $J_2$ as free parameters. The global solution over all datasets $T$ $>$ $T_{\text{N}}$   yield $J_1$ = 0.64(6) K and $J_2$ = 0.29(2) K. 

In a second step we analyzed the short-range correlations below $T_{\text{N}}$. We fitted the long- and short-range  contributions with two-dimensional Gaussians and Lorentzians, respectively. Two different Gaussian widths were used to account for the anisotropic instrument resolution of the long-range ordered Bragg peaks. In contrast only one width was used for the Lorentzians, accounting for the short-range order that substantially exceeds the instrumental resolution. After the Gaussian contributions were subtracted from the diffraction patterns below $T_{\text{N}}$, they were analyzed with Spinteract\cite{Paddison_2023}. In our analysis we find that identical $J_1$ and $J_2$ parameters as for the data above $T_{\text{N}}$ can be used, if the global scaling parameter is reduced by 50\%. This amounts to a magnetic moment value of 1-$\sqrt{0.5}$ = 29.3\%  that is fluctuating within the long-range ordered state.

\subsection*{Supplemental Note 3. Inelastic Neutron Scattering Analysis of Magnetic Field-Polarized Spin-Waves}
The field-polarized spin waves were recorded on CAMEA using an 11 T-vertical cryo-magnet with dilution insert. The experiment was performed at $T$ = 40 mK and $\mu_0H$ = 4, 5 and 6.5 T. In each scan the sample was rotated in 1 degree steps over a region of 120 degree between the structural (2, 0, 0) and (0, 2, 0) Bragg peaks, up to $|\vec{Q}|$ = 2.83~\AA$^{-1}$ in an energy transfer window from $E$ = 0 to 2 meV. We measured with setups $E_i$ = 5 meV and $2\theta$ = -44 and -79 degrees, with respective monitors $M$ = 500'000 and 250'000 (approx. 120 and 60 second/point). Each setup consisted of four individual scans in between which the detector position was shifted by 4 degrees and/or the energy by 0.13 meV to cover all blind spots between the analyzer units.

The data was reduced with the MJOLNIR \cite{Lass_2020} software package (version 1.3.1) The Currat-Axe spurions were masked using a circle of $\Delta Q$ = 0.07 \AA$^{-1}$ in diameter. One-dimensional cuts through the data were performed at various positions in reciprocal space and fitted to Gaussians to extract the spin-wave intensity, energy and width. An integration width $\Delta Q$ = 0.2 rlu and step size $\delta Q$ = 0.05 rlu perpendicular and along the cuts were used. We fixed the energy step size to $\delta E$ = 0.04 meV. In total 820 reciprocal space points were fitted.

The spin-wave spectra at the three magnetic fields were analyzed using the spin-1 Heisenberg Hamiltonian

\begin{equation}
H = J_1 \sum_{\langle i,j\rangle} \vec{S_i} \cdot \vec{S_j} + J_2 \sum_{\langle \langle i,j\rangle \rangle} \vec{S_i} \cdot \vec{S_j} - g\mu_B\mu_0H \sum_{i} S_i^{z}
\label{equ:Hamiltonian_SpinW}
\end{equation}

The Zeeman term (third term in Eq. \ref{equ:Hamiltonian_SpinW}) accounts for the magnetic anisotropy that is imposed by the magnetic field direction.
Using the Su(n)ny software package \cite{Sunny_2023} (version 0.6.0) and PyCrystalField \cite{PyCristalField_2021} (version 2.3.1), we consistently found that the linear Zeeman term is only approximate for the CEF scheme of HoInCu$_4$ (see Suppl. Fig. \ref{fig:g_factor}a). Thus, we simulated the corresponding field dependent g-factor $g$($H$) = $E$/$\mu_B\mu_0H$ of the lowest excited state associated to the field-split CEF triplet ground state and overplotted the field-polarized spin-wave maxima at (2, 0, 0) and (0, 0, 2) (see Fig. Suppl. \ref{fig:g_factor}b). Here $\mu_B$ is the Bohr magneton and $E$ the excitation energy into the lowest excited state. The results are in qualitative agreement with one another so that for the further analysis the values $g$ = 4.1(3), 3.8(2) and 3.6(2) were used for $\mu_0H$ = 4, 5 and 6.5 T, receptively. Discrepancies with the calculations were attributed to uncertainties in the CEF parameters, that the $J_1$ and $J_2$ were neglected in the simulation, and that dipole-dipole interactions were not considered (see Note 6). Thus, we note that more accurate results can be obtained if the interaction parameters are refined using random phase approximation (RPA) together with the CEF wave-functions in the field-polarized spin-wave analysis. This analysis will be published separately together with a software package containing the RPA code. We mention that the simulations in Suppl. Fig. \ref{fig:g_factor}a supports that no hybridization of the CEF ground state triplet and excited doublet takes place. The magnetic field dependence of the excitation branches reveals no change in slope around their intersection, which would be expected in a hybridized scenario. 

\begin{figure}
\centering
\includegraphics[width=0.8\linewidth,clip]{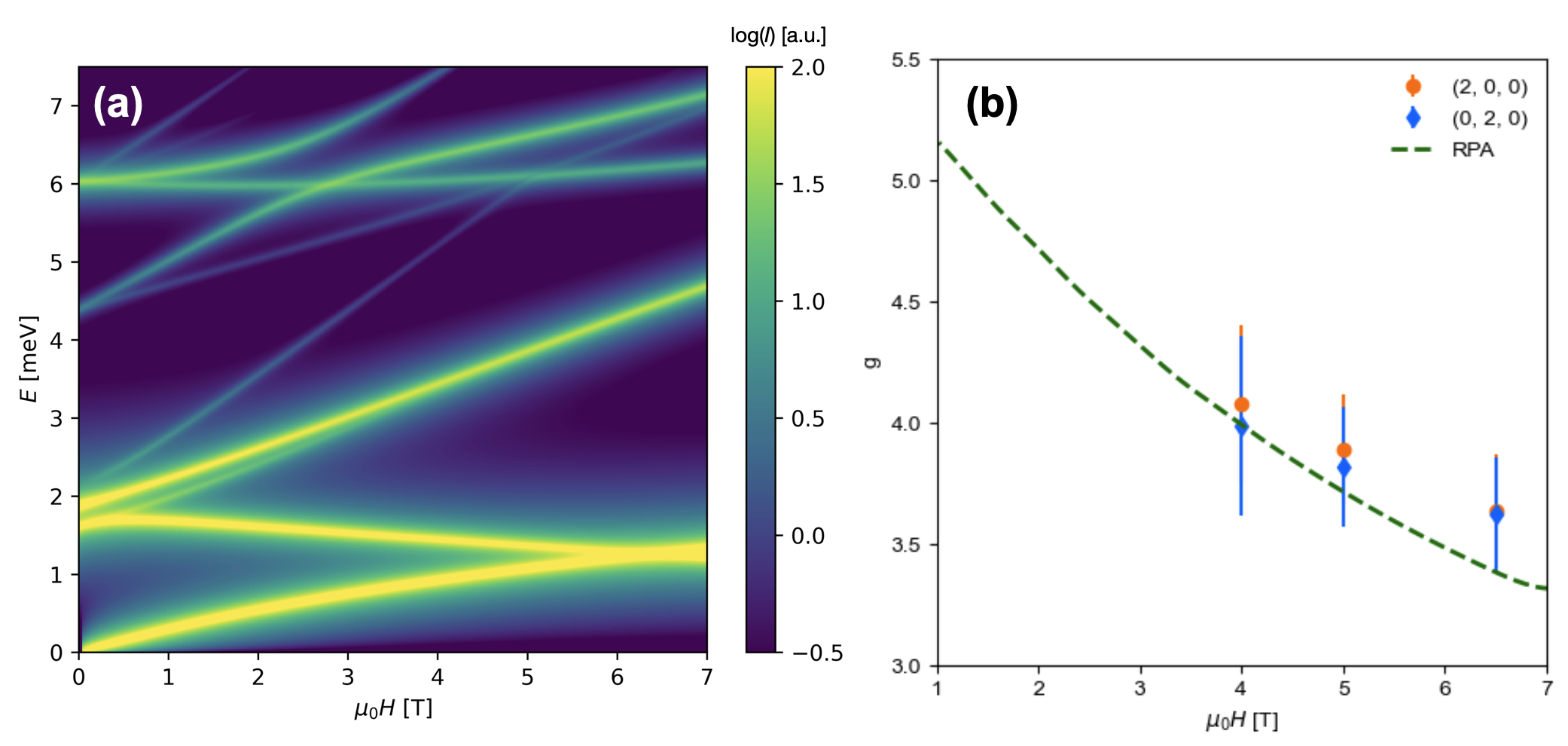}
\caption{\textbf{(a)} Modeled magnetic field dependence of the CEF scheme at $T$ = 2 K. The color bar represents the neutron intensity on a logarithmic scale. \textbf{(b)} Field dependent g-factor $g$($H$) = $E$/$\mu_B\mu_0H$ of the lowest excited state associated to the field-split CEF triplet ground state and overplotted with the field-polarized spin-wave maxima at (2, 0, 0) and (0, 0, 2).}
\label{fig:g_factor}
\end{figure}

The linear spin-wave analysis was performed with SpinW \cite{Toth_2015} (version 3.1). The magnetic structure was refined at each field using the interaction parameters obtained from the diffuse scattering data and the $g$-factors of the corresponding fields. These interaction parameters also served as starting parameters for the spin-wave refinements at the three fields, which yield the global solution $J_1$ = 0.66(3) and $J_2$ = 0.30(3) K. No other local minima were found.

\subsection*{Supplemental Note 4. Zero-Field Magnetic Excitation Spectrum}

The low-energy spin fluctuations of HoInCu$_4$ were studied at $T$ = 40 mK using CAMEA equipped with a 11T-vertical cryomagnet with dilution insert. The main dataset was acquired with a sample aligned perpendicular to the $c$-axis that was rotated in 1 degree steps over 100 degrees, covering the structural (2, 0, 0) and (0, 2, 0) Bragg peaks in an energy window up to $E$ = 0.9 meV. We used incident energies $E_i$ = [3.34, 4.04] meV at detector positions $2\theta$ = [-66, -62] degrees to achieve an elastic energy resolution of 100 $\mu$eV. A monitor $M$ = [875'000, 1'000'000] corresponding to $\tau$ = [197, 225] s/point was used. Each setup consisted of four individual scans in between which the detector position was shifted by 4 degrees and/or the energy by roughly 0.1 meV to cover all blind spots between the analyzer units. A background dataset was taken at $\mu_0H$ = 10 T for which the low-energy spectrum did not contain magnetic excitations, and was subtracted from the data. In order to obtain a full picture of the magnetic excitations we carried out a separate measurement, in which the sample was aligned perpendicular to the [1, -1, 0]-axis. Here, the sample was rotated in 1 degree steps over 130 degrees, covering the structural (0, 0, -1) and (1, 1, 0) positions in an energy window up to $E$ = 0.9 meV. We used incident energies $E_i$ = [3.34, 4.04] meV at detector positions $2\theta$ = [-46, -46] degrees to achieve an elastic energy resolution of 100 $\mu$eV. A monitor $M$ = [750'000, 1'000'000] corresponding to $\tau$ = [169, 225] s/point was used. Each setup consisted of four individual scans in between which the detector position was shifted by 4 degrees and/or the energy by roughly 0.1 meV to cover all blind spots between the analyzer units. A background dataset was taken at $\mu_0H$ = 7 T for which the low-energy spectrum did not contain magnetic excitations, and was subtracted from the data. Both datasets were reduced with the MJOLNIR \cite{Lass_2020} software package (version 1.3.1), with which we also masked the Currat-Axe spurions using a circle of $\Delta Q$ = 0.07 \AA$^{-1}$ in diameter. Supplemental Figure \ref{fig:Qplanes}a and b show constant energy cuts at energy transfers $E$ = 0.00(5), 0.20(5) and 0.35(5) meV in the ($H$, $K$, 0)- and ($H$, $H$, $L$)-plane, respectively, . 

\begin{figure}
\centering
\includegraphics[width=1\linewidth,clip]{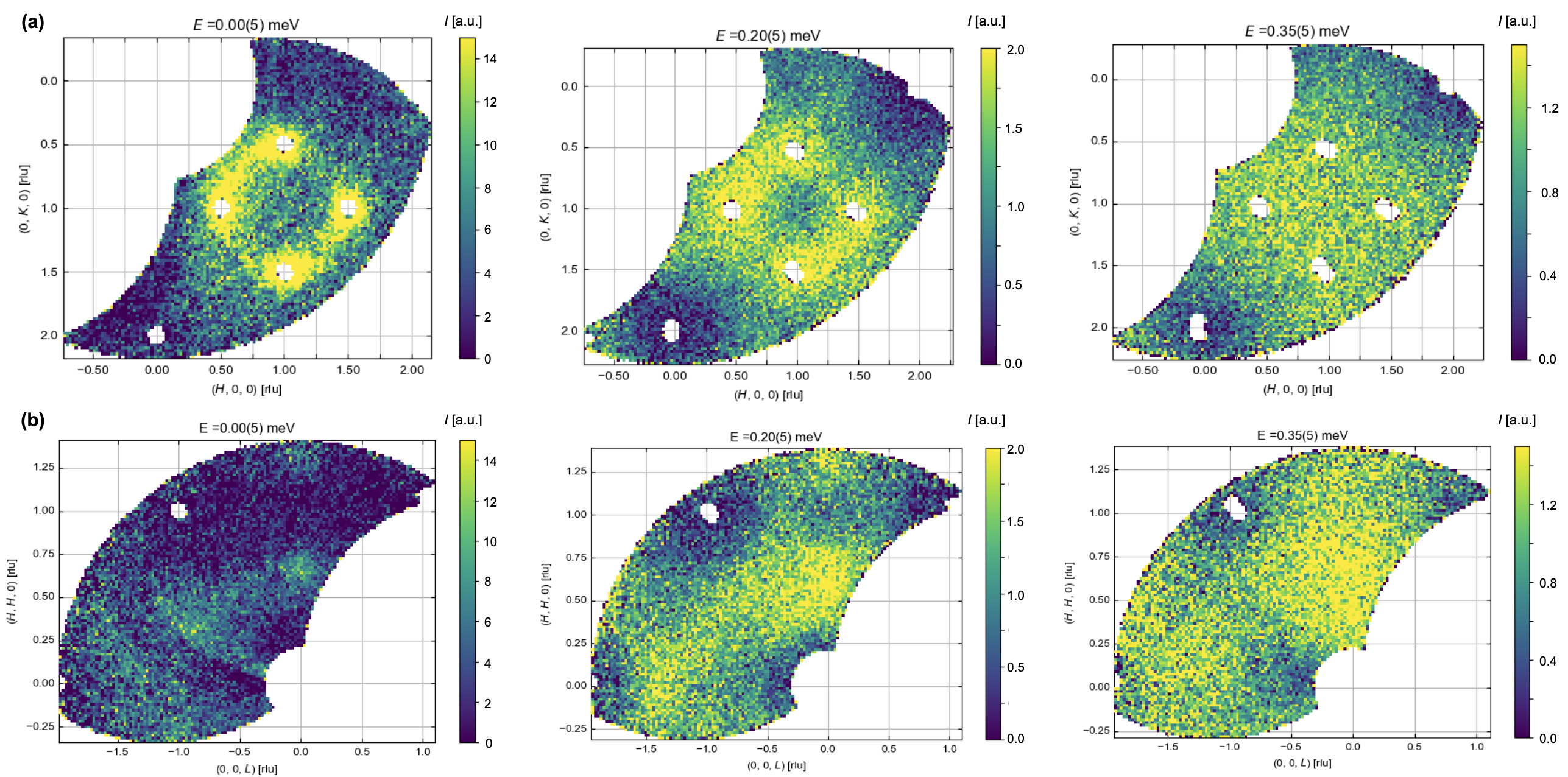}
\caption{Colour-coded background subtracted reciprocal space maps at $E$ = 0.00(5), 0.20(5) and 0.35(5) meV in the ($H$, $K$, 0)- and  ($H$, $H$, $L$)-planes in \textbf{(a)} and \textbf{(b)}, respectively. The colour bar represents the neutron intensity $I$.}
\label{fig:Qplanes}
\end{figure}

Linear spin-wave calculations were performed with the  SpinW \cite{Toth_2015} (version 3.1) and Su(n)ny \cite{Sunny_2023} (version 0.6.0) software packages. Supplementary Figure \ref{fig:sunny}a shows the magnetic excitation spectrum expected for a type-III AFM state using $J_1$ = 0.66 and $J_2$ = 0.30 K, the interaction parameters found in the magnetic-field polarized and paramagnetic states. The spectrum features two magnon branches below $E$ $<$ 0.2 meV that are associated to the two magnetic orbits. They originate from the ordering wavevector $\vec{q} = (1,\frac{1}{2}, 0)$ splitting the face-centered cubic lattice into independent Ho sublattices (see Fig. 1a of the main manuscript). Thus, a hypothetical half-ordered state in which only one sublattice is ordered while the other one does not exhibit long-range order is one way to account for the reduction of the ordered magnetic moment in HoInCu$_4$\cite{Anderson1950,Villain1959,Nakamura_1999,Oliver_2020}. In Suppl. Fig. \ref{fig:sunny}b we plot the excitation spectrum for a state in which we consider only one ordered face-centered cubic sublattice, revealing that the two branches merge into a single branch. We used a Lorentzian energy smearing of $\Delta E$ = 0.04 meV half-width half-maximum to simulate the magnetic exctiation spectra on CAMEA (see Suppl. Figs. \ref{fig:sunny}c and d), underlining that the two cases are indistinguishable in our experiment. 

\begin{figure}
\centering
\includegraphics[width=1\linewidth,clip]{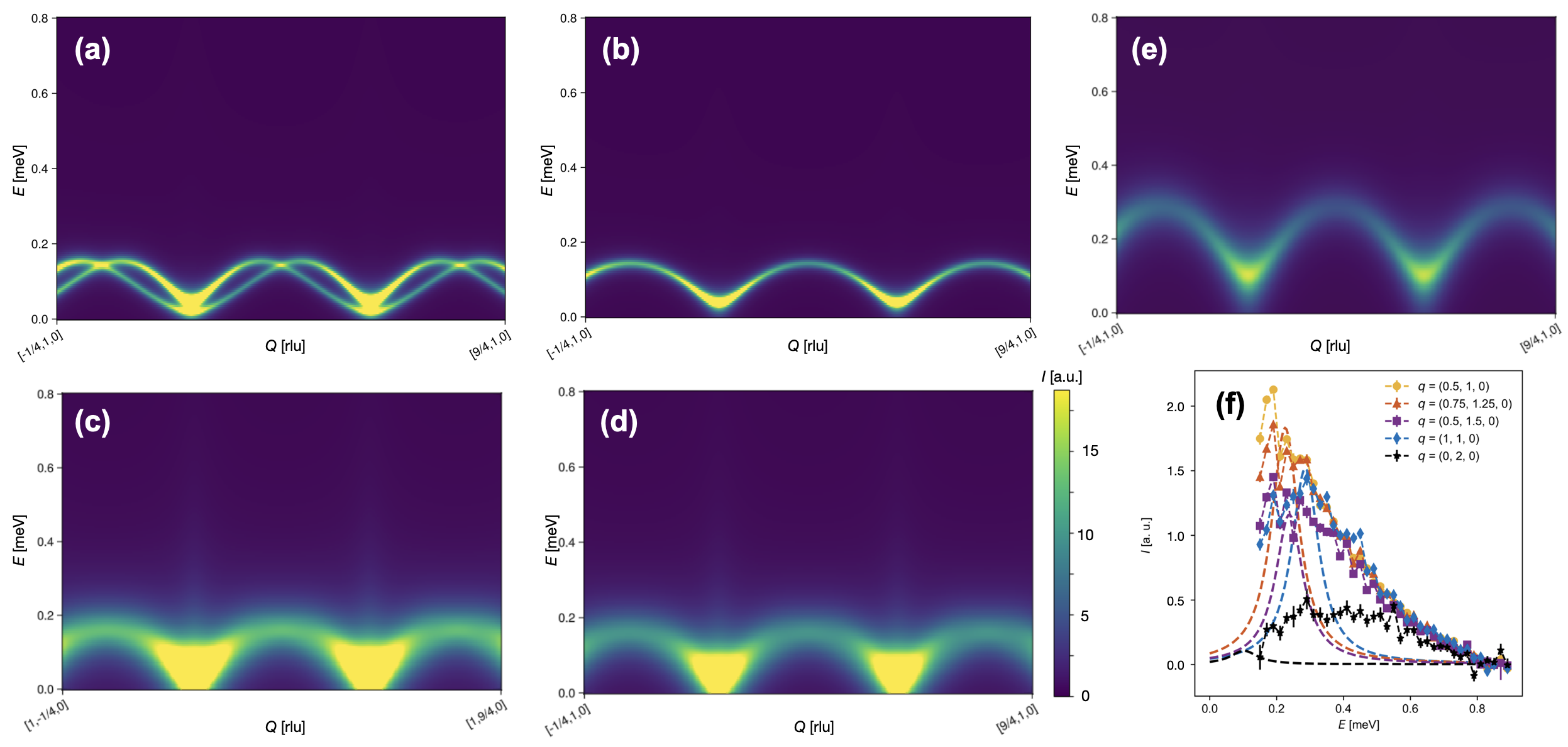}
\caption{Linear spin-wave simulation of a fully-ordered \textbf{(a)} and half-ordered \textbf{(b)} type-III AFM state using $J_1$ = 0.66 and $J_2$ = 0.30 K. The half-ordered state assumes that only one face-centered cubic sublattice is ordered. A Lorentzian energy smearing of $\Delta E$ = 0.04 meV half-width half-maximum was used to simulate the expected spectra of the fully- \textbf{(c)} and half-ordered \textbf{(d)} states on the CAMEA instrument. \textbf{(e)} Linear spin-wave simulation of a half-ordered type-III AFM state scaling $J_1$ and $J_2$ by 1.75. \textbf{(f)} One-dimensional line cuts through the experimental data using a circular integration width $\Delta Q$ = 0.2 rlu, overplotted with one-dimensional cuts through the simulation in panel \textbf{(d)}.}
\label{fig:sunny}
\end{figure}

We mention that in the case of a half-ordered Ho orbit, one would naively expect to observe coherent spin waves from the long-range ordered moments that are superimposed with broad excitations, originating from short-range correlated Ho moments. In contrast to the hypothesis, no signatures of coherent spin waves are visible in the inelastic neutron scattering spectra shown in Fig. 4b of the main manuscript. Moreover, the determined interaction parameters would have to be massively rescaled to qualitatively reproduce the energy maxima in Fig. 4b and d of the main manuscript. In order to test this, we modified the interaction parameters $J_1$ and $J_2$ keeping $J_2$/$J_1$ = 0.45 to qualitatively match the experimental observations. An acceptable overlap is found with a scaling factor of 1.75 using a half-ordered state where only one sublattice is ordered (see Figs. Suppl. \ref{fig:sunny}e and f). The one-dimensional energy cuts through the experimental data in Suppl. Fig. \ref{fig:sunny}f are overplotted with cuts through the simulation in Suppl. Fig. \ref{fig:sunny}e. They show that we can qualitatively reproduce the intensity maximum around $E$ = 0.3 meV, but that the magnetic excitations are much broader than what we would expect from the energy resolution CAMEA provides. Together with the unphysical scaling factor of 1.75 we conclude that the dynamical properties of HoInCu$_4$ originate from a more complex microscopic state. 

We also note that one scenario resulting in finite magnon life-times is a coupling between the two sublattices, triggering a decay of coherent spin waves. The lowest exchange interaction connecting the two sublattices is the third-nearest neighbour term $J_3$. However, refinements of the field-polarized excitation spectra including $J_3$ yield a minuscule, if finite, value $J_3$ = 0.02(2) K. Finite domain size effects are another possible mechanism that hinders the emergence of coherent spin waves, but the average magnetic domain size in HoInCu$_4$ is $\xi \approx$ 200 \AA~and the excitation life-time $\tau \approx$ 20 ps (see Figs. 2c and 4d of the main manuscript). Thus, a spin-wave velocity $v >$ 49 meV/\AA$^{-1}$ is required for domain effects to matter, which is at least an order of magnitude larger than what the refined interaction parameters predict.

\subsection*{Supplemental Note 5. Magnetic Field Dependence of the Ordered and Half-Ordered Type-III Antiferromagnetic States}

The magnetic field dependence of the long- and short-range AFM order was studied on the cold diffractometer DMC, PSI, which is equipped with a two-dimensional area detector spanning over $\Delta 2\theta$ = 132 and $\Delta \nu$ = $\pm$7 degrees in the horizontal and vertical scattering plane, respectively. We used a wavelength of $\lambda$ = 4.5189 \AA. The sample was measured in a 10 T-vertical magnet with dilution insert. Sample rotation scans at $T$ = 0.1 K with a step sizes of 0.1 degrees were used to either cover large area over 90 degrees or smaller areas of 40 degrees around specific magnetic Bragg peaks. The field evolution of the magnetic signal was determined at $\mu_0H$ = 0, 0.04, 0.125, 0.25, 0.375, 0.5, 0.625, 0.75, 1, 1.25, 1.5 ,2, 3, 4 and 5 T. A background was measured at $T$ = 50 K and subtracted from the data. The data reduction was performed with the DMCpy software package\cite{Lass2025} (version 1.0.1). One dimensional line cuts through the magnetic (1, 1.5, 0) Bragg peak are shown in Suppl. Fig. \ref{fig:DMC_field}. An integration width $\Delta Q$ = 0.1 rlu perpendicular to the cut was used.

The simulations of the field dependent magnetic ordering was carried out via the Su(n)ny software package \cite{Sunny_2023} (version 0.6.0). We minimized the energy of the magnetic Hamiltonian at different magnetic fields, using the interaction parameters obtained from the field-polarized state. The intensities corresponding to the type-III AFM domains are shown in Suppl. Fig. \ref{fig:DMC_field}b and overplotted with the normalized intensities obtained from the diffraction experiment. The Figure shows that linear spin-wave theory correctly predicts the critical field around $\mu_0H \approx$ 2.5 T, but fails to capture the nature of the magnetic state between $\mu_0H \approx$ 1 and 2.5 T.

\begin{figure}
\centering
\includegraphics[width=0.8\linewidth,clip]{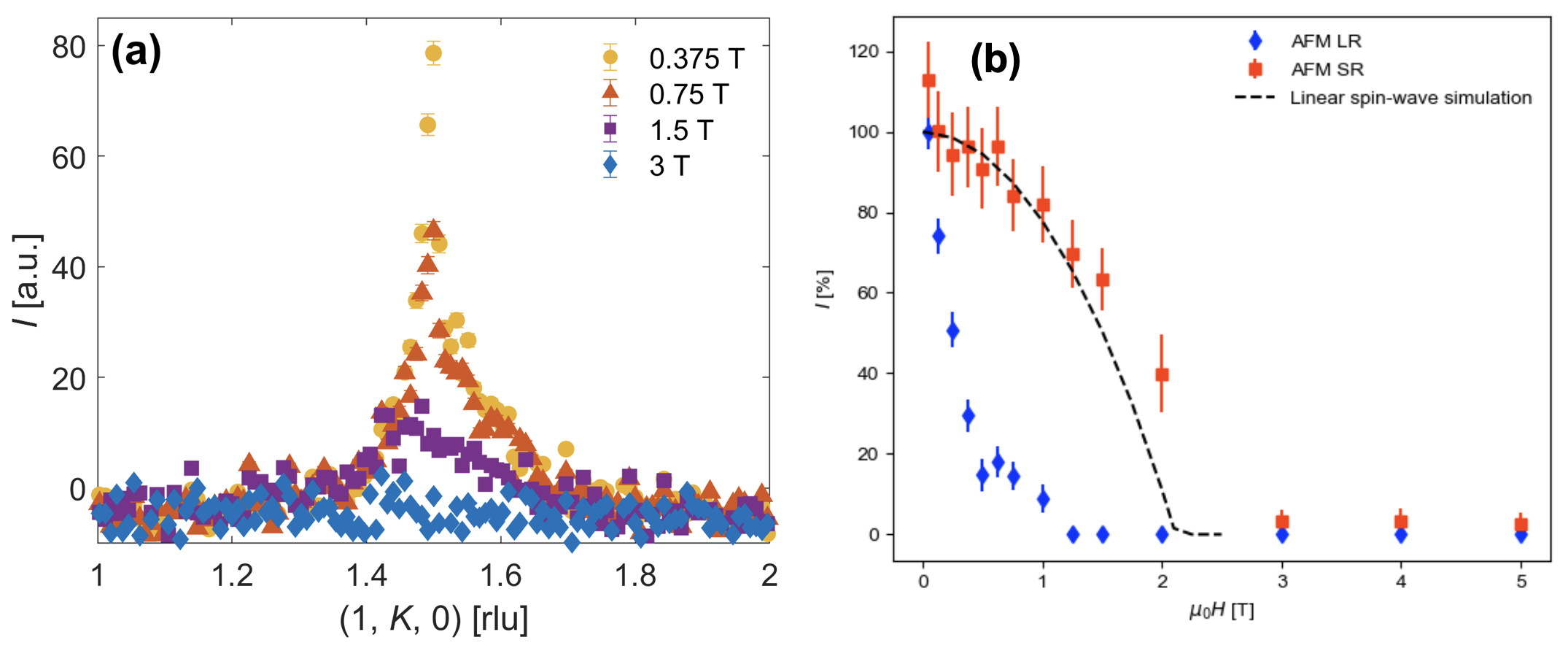}
\caption{\textbf{(a)} One dimensional line cuts through the magnetic (1, 1.5, 0) Bragg peaks using an integration width $\Delta Q$ = 0.1 rlu perpendicular to the cut. The data were taken at DMC at $T$ = 0.1 K. \textbf{(b)} Normalized field dependence of the antiferromagnetic long- (AFM LR) and short-range (AFM SR) order, overplotted with the field dependence of the linear spin-wave simulation.}
\label{fig:DMC_field}
\end{figure}

\subsection*{Supplemental Note 6. Effect of Dipole-Dipole Interaction}

The dipole-dipole interaction among magnetic ions can be sizable for large magnetic moments, and can become particularly relevant in materials with low N\'eel temperatures \cite{Bramwell2001}. The interaction strength can be estimated by
\begin{equation}
J_d = \frac{5}{3}\frac{\mu_0}{4\pi}\frac{\mu^2}{r^3}
\label{equ:dipole}
\end{equation}
with $\mu_0$ = 4$\pi$10$^{-7}$ N/A$^2$, $\mu$ the magnetic moment and $r$ the distance between moments. The nearest-neighbour distance in HoInCu$_4$ amounts to $r$ = 5.076 \AA. Using the moment expected by the ground state triplet $\mu_{CEF}$ = 4.58$\mu_B$ we find  $J_d$ = 170 mK, which is much smaller than $T_N$ = 760 mK and of the order of $J_2$/2. However, if we consider that only $\mu$ = 3.23$\mu_B$ is ordered a dipolar exchange coupling $J_d$ = 80 mK ($\sim J_2$/4) is obtained. This suggests that dipolar interaction have only a minor impact on the magnetic excitation spectrum of this material.

We investigated the dipole-dipole effect on the dynamic spectra by comparing simulations in which no dipolar interactions were assumed with simulations including long-range dipolar interactions. In Suppl. Fig. \ref{fig:dipole} we depict linear spin-wave simulations ($J_1$ = 0.66 K and $J_2$ = 0.3 K) at $\mu_0H$ = 5 (panels a and b) and 0 T (panels c and d) with and without long-range dipole interactions. The simulations were carried out using the Su(n)ny \cite{Sunny_2023} (version 0.6.0) software package, in which an Ewald summation over the dipolar interactions can be performed. The solid blue lines are the calculated magnon bands which are color-coded with the expected neutron intensity. We applied a Lorentzian smearing  $\Delta E$ = 0.08 and 0.04 meV half width half maximum for $\mu_0H$ = 5 and 0 T, respectively, mimicking the the instrument resolution. The dashed black lines in panels b  and d indicate the magnon bands without long-range dipole interactions.

\begin{figure}
\centering
\includegraphics[width=0.8\linewidth,clip]{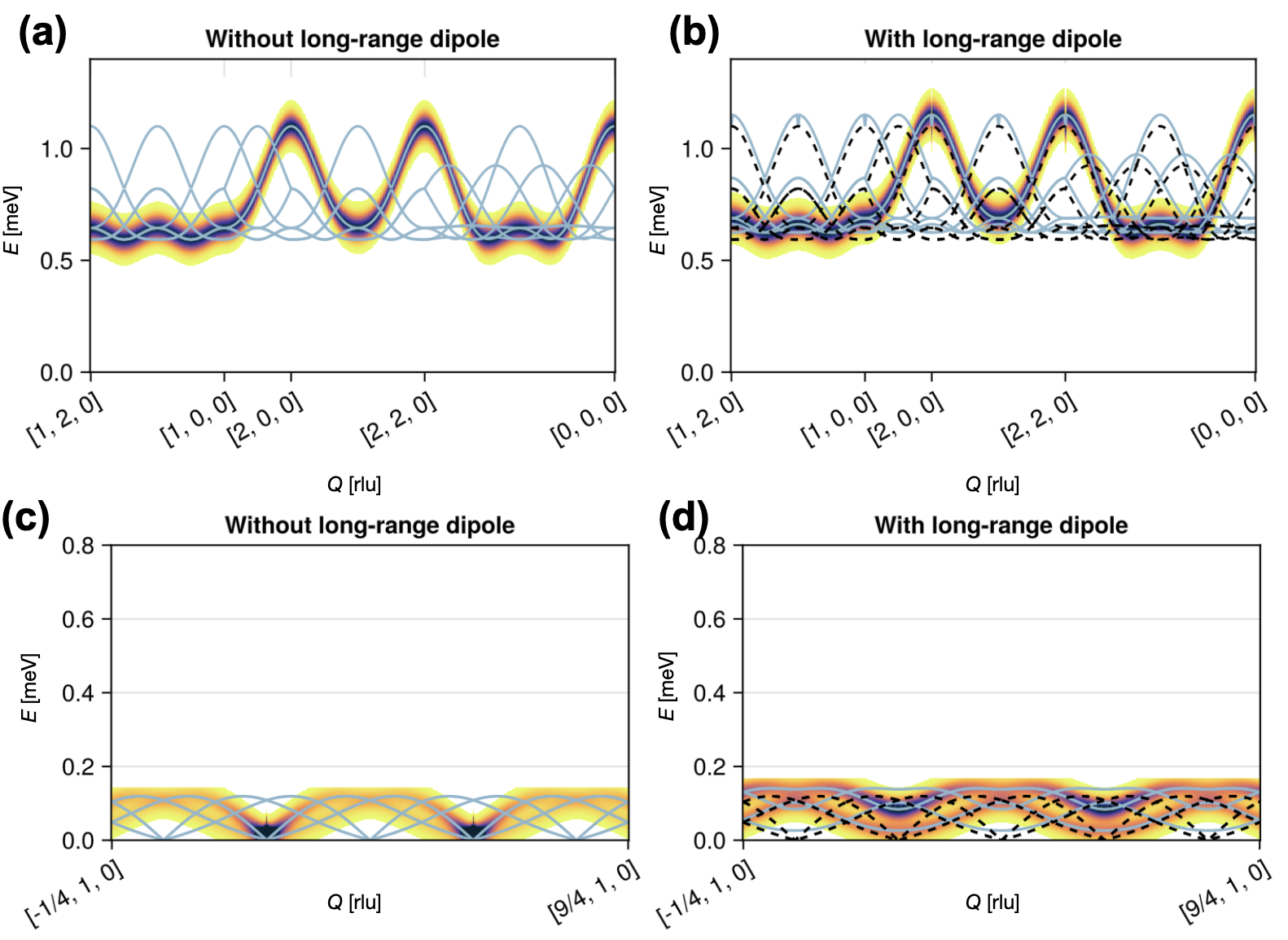}
\caption{Linear spin-wave simulations using the mean of the refined parameters $J_1$ = 0.66(3) K and $J_2$ = 0.30(3) K at $\mu_0H$ = 5 (\textbf{(a)} and \textbf{(b)}) and 0 T (\textbf{(c)} and \textbf{(d)}) with and without long-range dipole interactions. The solid blue lines are the calculated magnon bands which are color-coded with the expected neutron intensity. We applied a Lorentzian smearing  $\Delta E$ = 0.08 and 0.04 meV half width half maximum for $\mu_0H$ = 5 and 0 T, respectively, mimicking the the instrument resolution. The dashed black lines in \textbf{(b)}  and \textbf{(d)} indicate the magnon bands without long-range dipole interactions.}
\label{fig:dipole}
\end{figure}

The comparison confirms that dipole interactions have only a minor effect in HoInCu$_4$. The $\mu_0H$ = 5 T results show that the dipole interactions shift the magnon by < 50 $\mu$eV to higher energy transfer, an effect that is within the reported errors of $J_1$, $J_2$ and $g$. In fact, our simulations in Suppl. Fig. \ref{fig:g_factor}b suggest that the dipole-dipole effect is mainly absorbed in the uncertainty of $g$. In the Figure we compare the simulated g-factor of the lowest excited state associated to the field-split crystal field triplet ground state with the field polarized spin-wave maxima at (2, 0, 0) and (0, 0, 2). We find that the experimental results are shifted to slightly higher energy transfers when compared to the simulation, which is in line with the shift reported in Suppl. Fig. \ref{fig:dipole}b. However we mention that the discrepancies in  Suppl. Fig. \ref{fig:g_factor}b can also arise from uncertainties in the crystal-field parameters, and to the fact that the $J_1$ and $J_2$ interactions were not considered in the simulated g-factor field dependence. The spin-wave simulations at $\mu_0H$ = 0 T predict that the excitation spectrum is gapped by $\sim$40 $\mu$eV and that the two magnon bands split by $\sim$50 $\mu$eV at the magnetic Bragg peak position, so that the spin wave boundary is shifted from $E$ = 0.12 to 0.14 meV when compared to the simulation neglecting dipole interactions (see Suppl. Fig. \ref{fig:g_factor}c and d). Thus, linear spin-wave theory including dipole interactions is also unable to capture the low-energy spin dynamics at zero field.

}

\end{document}